\documentstyle[seceq,mbf,wrapft]{ptptex}

\def\susy23{S{\small USY23}}

\pubinfo{Vol.~XX, No.~X, October 1999}  

\title{
Automatic Computation of Cross Sections in HEP\footnote{Presented by
F.Yuasa at the ICCP5 conference, Kanazawa, October 1999}
}
\subtitle{Status of GRACE System}    

\author{
F.{\sc YUASA}\(^{a} \),
J.{\sc FUJIMOTO}\(^{a} \),  T.{\sc ISHIKAWA}\(^{a}
\),  M.{\sc JIMBO}\(^{b} \),  T.{\sc KANEKO}\(^{c} \),  K.{\sc KATO}\(^{d} \), 
S.{\sc KAWABATA}\(^{a} \),  T.{\sc KON}\(^{e} \), Y.{\sc KURIHARA}\(^{a} \),
M.{\sc KURODA}\(^{c} \),  N.{\sc NAKAZAWA}\(^{d} \), Y.{\sc
SHIMIZU}\(^{a} \),  H.{\sc TANAKA}\(^{f} \)}

\inst{
a)High Energy Accelerator Research Organization (KEK),
1-1 OHO, Tsukuba, Ibaraki 305-0801, Japan\\
b)Tokyo Management College, 625-1 Futamata, Ichikawa, Chiba
272-0001,Japan\\
c)Meiji-Gakuin University, Kamikurata 1518, Totsuka, Yokohama 244-0816,
Japan\\
d)Kogakuin University, Nishi-Shinjuku 1-24, Shinjuku, Tokyo 163-8677, Japan\\
e)Seikei University, Musashino, Tokyo 180-8633, Japan\\
f)Rikkyo University, 3-34-1 Nishi-Ikebukuro, Toshima, Tokyo 171-8501, Japan
}

\recdate{
October 11, 1999}

\abst{
For the study of reactions in High Energy Physics (HEP)
automatic computation systems have been developed and are
widely used nowadays.
GRACE is one of such systems and
it has achieved much success in analyzing experimental data.
Since we deal with the cross section whose value can be
given by calculating hundreds of Feynman diagrams, we
manage the large scale calculation, so that
effective symbolic manipulation, the treat of
singularity in the numerical integration are required.
The talk will describe the software design of GRACE system and 
computational techniques in the GRACE.
}

\begin{document}

\maketitle


\makeatletter
\if 0\@prtstyle
\def\asp{.3em} \def\bsp{.26em}
\else
\def\asp{.3em} \def\bsp{.3em}
\fi \makeatother

\section{Introduction}

High energy experimental physics has produced
excellent results thanks to the progress of the detectors and the
development of the high energy accelerators with high luminosity.
Accordingly, the accurate theoretical computation is required
to compare experimental results with theoretical predictions. On 
the other hand, as the beam energy becomes higher, 
there appear the physics
processes with many final state particles. As a matter of
course the number of relevant Feynman diagrams becomes huge.
This means that Feynman amplitude calculation 
is practically impossible when 
one calculates cross sections by hands. 

\par
Since the perturbative calculation in quantum field theory is
well-defined, it can be realized as an automatic computation system. 
There appeared several systems, for instance, 
{\tt CompHEP}\cite{Ref:COMPHEP} and {\tt FeynArt/FeynCalc}\cite{Ref:FEYN},
in HEP and nowadays
the automatic computation of Feynman amplitudes becomes common.

We have developed an automatic computation system named GRACE.
\cite{Ref:Graceman}
In section 2, we describe the feature of GRACE system
as a concrete example of such system. In section 3,
several physical achievements by GRACE are briefly explained.
In section 4, we discuss the techniques in GRACE from the view point
of computing. In the final section, we summarize the problems and the
future plans.

\section{GRACE System}

\subsection{Tree-Level}

Using GRACE, the computation proceeds in following steps:
\begin{enumerate}
\item{define a physics process with a model
and generate Feynman diagrams relevant 
to the process,\cite{Ref:GRC}}
\item{generate FORTRAN code for the numerical computation referring to
helicity amplitude library (CHANEL)\cite{Ref:CHANEL}}
\item{compute the cross section
by Monte Carlo integration package (BASES),\cite{Ref:BASES51}}
\item{simulate the events by event generation package (SPRING).\cite{Ref:BASES51}}
\end{enumerate}
\par
GRACE system consists of several component-programs corresponding to
each step.  
In the first step program named {\tt grc} generates diagrams. 
It can generate diagrams at any order of perturbation.
All the resultant information on 
the diagram topology, particles and
vertices is stored in the list-formated file.
In the next step program named {\tt grcfort} generates FORTRAN code
including interfaces to CHANEL and BASES. 
In the third step, two programs {\tt integ} and {\tt spring} are produced.
Before proceeding to the third step we have to care about kinematics for the
phase space integration using BASES and SPRING.
Running {\tt integ}, the cross section 
is computed numerically. Since the distribution of the integrand is
known in course of the integration,
unweighted events are produced by {\tt spring} in the last step.

\par
In GRACE, 
several physics model files and kinematics routines are necessary
and standard one's are built-in. 
User-defined model file or user's kinematics
file can replace built-in files.
GRACE also includes utility programs such as diagram-drawer,
{\tt gracefig}, and so on.  

\subsection{One-Loop Level}

The important extension of GRACE is to compute the loop
amplitudes automatically. 
As for the one-loop order, automatic computation of  the processes with 
two final state particles ($2 \rightarrow 2$) is almost completed
and
some $2 \rightarrow 3$ processes are computed by GRACE.
However,
it is currently restricted to the electro-weak interaction.

\par
The steps for the one-loop computation is almost same as those for the
tree-level. 
However, extra steps are included due to new features needed for
treating loop diagrams. 
We introduced the symbolic manipulation into GRACE to treat them and it
processes the following steps:
\begin{enumerate}
\item[(i)] {take the trace if there is a fermion line,}
\item[(ii)] {introduce Feynman parameters integration method,}
\item[(iii)] {shift a loop momentum appropriately,}
\item[(iv)] {drop odd orders of $l$,}
\item[(v)] {replace even orders of $l$ as $l^{\mu}l^{\nu}
\rightarrow g^{\mu\nu}l^2/n$ and so on, }
\item[(vi)] {contract vertex indices inside the loop in
$n$-dimension. ($n = 4 - 2\epsilon$) }
\end{enumerate}

In order to regularize the ultra-violet divergence and the
infrared divergence, 
the dimensional regularization 
and the 
fictitious photon mass parameter $\lambda$ are introduced into GRACE,
respectively.
\par
For the loop process program {\tt grc} generates both relevant tree
diagrams and loop diagrams at once. 
{\tt grc} also generates diagrams with vertex and propagator counter terms.
After generating diagrams, 
following steps are needed as a part of step 2. in the tree-level procedure:
\begin{itemize}
\item{Generate source program for the symbolic manipulation system,
REDUCE\cite{Ref:REDUCE} or FORM\cite{Ref:FORM}. Here, the product $T^{loop}T^{tree\dagger}$ are
written down in a symbolic code for each pairs of an one-loop
amplitude and a tree amplitude.}
\item{Invoke REDUCE or FORM and get FORTRAN source code.
}
\end{itemize}
After we get FORTRAN source code, the cross sections can be calculated
with the loop library and the counter term library to
deal with Feynman integral and counter terms, respectively.

\subsection{How to check the results}
The important and serious issue in automatic computation is how
we can check the results.  
For the tree-level computation GRACE can check the gauge invariance of the
results by using unitary and covariant gauge. 
This does not work in the loop amplitude since the structure of the numerator
becomes complex. 
For one-loop level, we newly implemented the non-linear gauge (NLG)
\cite{Ref:NLG} for gauge invariance
check.  
We take a generalized NLG fixing condition for SM:
\begin{equation}
\begin{array}{rl}
{\cal L}_{\rm GF}=& \displaystyle{
-\frac{1}{\xi_W}\left|(\partial_{\mu}-ie\tilde{\alpha}A_{\mu}
-ig\cos\theta_W\tilde{\beta}Z_{\mu})W^{+\mu}
+\xi_W\frac{g}{2}(v+\tilde{\delta}H+i\tilde{\kappa}\chi_3)\chi^{+}\right|^2
}\\
{ } & { } \\
{ } & \displaystyle{
-\frac{1}{2\xi_Z}\left(\partial_{\mu}Z^{\mu}
+\xi_Z\frac{g}{2\cos\theta_W}(v+\tilde{\epsilon}H)\chi_3\right)^2
\quad -\frac{1}{2\xi_A}(\partial_{\mu}A^{\mu})^2}
\end{array}
\end{equation}

Using NLG gauge invariance, check has been performed for several
one-loop processes. 

\par
Besides gauge invariance check we can check the self-consistency of
the loop-implementation to GRACE by changing parameters $1/\epsilon$ or
$\lambda$. 
Comparison among the results derived from independent automatic computation
systems shows very good agreements.\cite{Ref:NLGGRACE}

\subsection{Beyond the Standard Model}

GRACE has been extended to compute the physics processes not only in
the Standard Model but also the Minimal Supersymmetric extension of the
Standard Model(MSSM). Here the treatment of the
Majorana particles and the fermion number clashing vertices
is introduced.\cite{Ref:GRACESUSY}
The MSSM includes 85 particles and 3,764 vertices so that
the automatic calculation becomes indispensable for the calculation
of SUSY processes.
Recently the complete first-order radiative
corrections to the process $H^+ \rightarrow t \bar{b}$
has been calculated using 1-loop system(GRACE/SUSY/1LOOP).\cite{Ref:HTB}

\section{Physics Achievements}

So far GRACE system has been proved to be very powerful in computing
many complicated processes with many final state particles. 
In the tree-level $2 \rightarrow 3$ (over 50 processes), $2
\rightarrow 4$ (about 100 processes), $2
\rightarrow 5$ ($e^+e^- \rightarrow e^-\bar{\nu_{e}} u \bar{d}
\gamma$\cite{Ref:ENEUDA}), and $2 \rightarrow 6$ ($e^+e^- \rightarrow b \bar{b} u
\bar{d} \mu \bar{\nu_{\mu}}$ \cite{Ref:BBUDMNM}) were computed.
Among them it is noteworthy that 76 $e^+e^- \rightarrow$ 4-fermion processes
which take place in the energy region above $W^+W^-$ pair threshold
are computed for LEP-II experiment. 
These 76 4-fermion programs are combined and made them up to the
program named {\tt grc4f}. \cite{Ref:GRC4F}
The one-loop level processes such as $e^+e^- \rightarrow Z^0H$, $e^+e^-
\rightarrow t \bar{t}$, $\gamma \gamma \rightarrow W^+W^-$,
and $e^+e^- \rightarrow W^+\mu \bar{\nu_{\mu}}$
 were computed.
\cite{Ref:ONELOOP}
\par
To provide a practical generator for SUSY processes at LEP-II and
a linear collider, we developed {\tt susy23}
which contains 23 processes for $e^+ e^-$ physics.
\cite{Ref:SUSY23}
GRACE/SUSY can be extended to include R-parity violating interactions
so that the process $e^+e^- \rightarrow \nu \bar{\nu_{e}} d \bar{d}$ in LEP-II
was calculated.\cite{Ref:RPARITY}
\par
Recently GRACE has been applied to a process at HERA
experiments(GRAPE).\cite{Ref:GRAPE}
It was the lepton pair production, both in the elastic
and the deep inelastic scattering regions, $ep \rightarrow epl\bar{l}$
and $ep \rightarrow el \bar{l}X$. Some routines for the proton form
factors and the structure functions have been implemented.

\section{Issues Related to Computation}

The first issue is the computation time.
As the physics process treated by GRACE becomes  more complicated, the
computation time becomes so longer that it 
restricts applications of the automatic computation.
Among the computing steps, the multi-dimensional Monte Carlo
integration consumes a lot of computing time. 
The requirements for reducing the execution time, we 
have successfully developed the first version of parallelized GRACE by
using message passing library, PVM or MPI.\cite{Ref:PVMMPI}
This version of parallelized
GRACE reduced the computation time drastically.\cite{Ref:PVMGRACE} 

We measured the reduction rate with
Fujitsu AP3000\footnote{The AP3000 system consists of UltraSPARC -II
300MHz processors connected via AP-Net. AP-Net is a two-dimensional
torus network and provides 200MB/s bandwidth per port.} by using up to
16 processors for the physics  
process $e^+e^- \rightarrow b\bar{b}u\bar{d}\bar{\nu_{\mu}} \mu$.
The speed up with 16 processors was 15.82.
\par
The second issue is related to a treatment of singularities appearing in the
multi-dimensional Monte Carlo integration.
In the framework of GRACE kinematics routine used in the phase space
integration is not generated automatically.
The aim of kinematics routine is to find an appropriate transformation
between integral variables and kinematics variables to keep away from
singular behavior of amplitudes in the integral region. However, we
don't know yet an algorithm good enough to deal with general singularities. To
find better algorithm for singular integrals is thus another big issue and it belongs to the
field of numerical integration on computer.\cite{Ref:DICE}

\par
The third issue is the symbolic manipulation.
\begin{itemize}
\item When we write REDUCE code naively,
the short of memory often happens during processing the
manipulation. In order to get rid of this, many
complicated and sophisticated techniques have been
introduced. However 
in manipulation for 1-loop of $2 \rightarrow 4$ process,
it is almost impossible to perform the symbolic 
manipulation by REDUCE.
FORM system can handle the lengthy code
because FORM needs less memory and computing time
than REDUCE.
\item When we perform the symbolic manipulation using
REDUCE and FORM, the long expressions(more than 10,000 lines)
are obtained. Then very long computing time is required in
numerical integration. The effective optimization should
be considered. 
The symbolical optimization can be possible through pattern matching
because Feynman diagrams contain many common parts.
\end{itemize}

\par
The last issue is the high precision calculation in GRACE. It is
sometime required from the following reasons:
\begin{itemize}
\item The all vertices of SM and MSSM had been checked by
gauge invariance in the quadruple precision.
A possible bug might happen in the kinematics code.
It can be detected by the gauge invariance check.
However, if it affects minor contribution, the double
precision is not enough to manifest the bug.
\item Due to several kinematical reasons the accuracy of numerical
values may be lost. 
For instance, the two photon process has drastic
numerical cancellation between Feynman diagrams. It may give
a wrong result if one uses the double precision ($10^{-15}$).
\item In the cross section formula, there can be a term
$(W/m_{e})^2$ which is $ \sim 10^{+14}$ in TeV collisions
where $m_{e}$ is the electron mass and $W$ is C.M. energy.
This indicates that double precision is no more enough
in HEP computation.
In loop calculation one will loose some accuracy
 in the evaluation of loop integrals in addition to
 kinematics. In some case the calculation with
full quadruple precision will be required. 
\end{itemize}

\section{Summary}

We have developed GRACE, an automatic computation system in HEP,
and it has made much success for the study of physics process
in tree and one-loop order.
In the course of study, we have confronted problems of large scale computations
that prevent the application of the system.
The restriction can be removed by the hardware development
and software engineering.
Among them, we have discussed here the parallel
computation, the singular integral in numerical computation, 
the symbolic manipulation and the floating number with high
precision.
Armored with those items, we can obtain the theoretical predictions
for a larger class of physical processes in the
current and future accelerator projects.

\section*{Acknowledgements}

The authors would like to thank the local organizing committee of ICCP5
for the excellent organization.
This work is supported by the Supercomputer Project No.49(FY1999)
of High Energy Research Organization (KEK) and in part by the
Ministry of Education, Science and Culture, Japan under the
Grant-in-Aid for Scientific Research Program No.11440083, No.10640282,
No.10680366 and No.10640285.

\end{document}